\documentclass[pre,twocolumn,superscriptaddress,floatfix]{revtex4}
\usepackage[latin1]{inputenc}
\usepackage{graphicx}
\usepackage{amsmath}
\usepackage{amssymb}
\usepackage{color}

\begin{document}

\title{Flow past superhydrophobic surfaces with cosine variation in local slip length}

\author{Evgeny S. Asmolov}
\affiliation{A.N.~Frumkin Institute of Physical
Chemistry and Electrochemistry, Russian Academy of Sciences, 31
Leninsky Prospect, 119991 Moscow, Russia}
\affiliation{Central Aero-Hydrodynamic Institute, 140180
Zhukovsky, Moscow region,  Russia}
\affiliation{Institute of Mechanics, M. V. Lomonosov Moscow State University, 119991 Moscow,
Russia}

\author{Sebastian Schmieschek}
\affiliation{Institute for Computational Physics, University of Stuttgart, Pfaffenwaldring 27, D-70569 Stuttgart, Germany}
\affiliation{Department of Applied Physics, Eindhoven University of Technology, P.O. Box 513, 5600 MB Eindhoven, The Netherlands}

\author{Jens Harting}
\affiliation{Department of Applied Physics, Eindhoven University of Technology, P.O. Box 513, 5600 MB Eindhoven, The Netherlands}
\affiliation{Institute for Computational Physics, University of Stuttgart, Pfaffenwaldring 27, D-70569 Stuttgart, Germany}
\author{Olga I. Vinogradova}
\affiliation{A.N.~Frumkin Institute of Physical
Chemistry and Electrochemistry, Russian Academy of Sciences, 31
Leninsky Prospect, 119991 Moscow, Russia}
\affiliation{Department of Physics, M.V.~Lomonosov Moscow State University, 119991 Moscow, Russia }
\affiliation{DWI, RWTH Aachen, Forckenbeckstr. 50, 52056 Aachen, Germany}
\date{\today}

\begin{abstract} Anisotropic super-hydrophobic surfaces have the
potential to greatly reduce drag and enhance mixing phenomena in
microfluidic devices. Recent work has focused mostly on cases of
super-hydrophobic stripes. Here, we analyze a relevant situation of
cosine variation of the local slip length. We derive approximate formulae
for maximal (longitudinal) and minimal (transverse) directional
effective slip lengths that are in good agreement with the exact
numerical solution and lattice-Bolzmann simulations for any surface
slip fraction. The cosine texture can provide a very large effective
(forward) slip, but it was found to be less efficient in generating a
transverse flow as compared to super-hydrophobic stripes.
\end{abstract}

\pacs {47.11.-j, 83.50.Rp,  47.61.-k}
\maketitle

\section{Introduction}
The design and fabrication of textured superhydrophobic surfaces have
received much attention in recent years. If the recessed regions of
the texture are filled with gas (the Cassie state), roughness can
produce remarkable liquid mobility, dramatically lowering the ability
of drops to stick~\cite{quere.d:2008}. These surfaces are known to be
self-cleaning and show low adhesive forces. In addition to the
self-cleaning effect, they also exhibit drag reduction for fluid
flow. Thus, they are of importantance in the context of transport
phenomena and fluid dynamics as
well~\cite{bocquet2007,vinogradova.oi:2011,rothstein.jp:2010}. Many
sea animals e.g. shark and other fish are known to possess
superhydrophobic skin~\cite{bhushan.b:2011}. Also many artificial
textures have been designed to increase drag reduction
efficiency~\cite{vinogradova.oi:2012}.

This drag reduction is associated with the liquid slippage past solid
surfaces. This slippage occurs at smooth hydrophobic surfaces and can
be described by the boundary
condition~\cite{vinogradova.oi:1999,bocquet2007,lauga2005},
$u_{\rm slip} = b \,\partial u / \partial z$,
where $u_{\rm slip}$ is the (tangential) slip
velocity at the wall, $\partial u / \partial z$ the local shear rate,
and $b$ the slip length. A mechanism for hydrophobic slippage involves
a lubricating gas layer of thickness $\delta$ with viscosity $\mu_g$
much smaller than that of the liquid
$\mu$~\cite{vinogradova.oi:1995a}, so that $b \simeq \delta (\mu/\mu_g
- 1) \simeq 50
\delta$~\cite{vinogradova.oi:1995a,andrienko.d:2003}. However, at
smooth flat hydrophobic surfaces $\delta$ is small, so that $b$ cannot
exceed a few tens of
nm~\cite{vinogradova.oi:2003,vinogradova.oi:2009,charlaix.e:2005,joly.l:2006}. In
case of superhydrophobic surfaces, the situation can change
dramatically, and slip lengths up to tens of $\mu$m may be obtained
over a thick gas layer stabilized with a rough
texture~\cite{choi.ch:2006,joseph.p:2006}.

To quantify the flow past heterogeneous surfaces it is convenient to apply the concept of an effective slip boundary condition at the imaginary smooth homogeneous,
but generally anisotropic surface \cite{vinogradova.oi:2011,Kamrin_etal:2010}.
Such an effective condition mimics the actual one along the true heterogeneous surface, and fully characterizes the real flow.
The quantitative understanding of the effective slip length of the superhydrophobic surface, $\mathbf{b}_{\mathrm{%
eff}}$, is still challenging since the composite nature of the texture in addition to liquid-gas areas requires
regions of lower local slip (or no slip) in direct contact with the liquid. For an anisotropic texture, the effective slip generally
depends on the direction of the flow and is a tensor, $\mathbf{b}_{\mathrm{%
eff}}\equiv \{b_{ij}^{\mathrm{eff}}\}$ represented by a symmetric, positive
definite $2\times 2$ matrix~\cite{Bazant08}
\begin{equation}
\mathbf{b}_{\mathrm{eff}}=\mathbf{S}_{\theta }\left(
\begin{array}{cc}
b_{\mathrm{eff}}^{\parallel } & 0 \\
0 & b_{\mathrm{eff}}^{\perp }%
\end{array}%
\right) \mathbf{S}_{-\theta },  \label{beff_def1}
\end{equation}%
diagonalized by a rotation
\begin{equation*}
\mathbf{S}_{\theta }=\left(
\begin{array}{cc}
\cos \theta & \sin \theta \\
-\sin \theta & \cos \theta%
\end{array}%
\right) .
\end{equation*}%
Eq.(\ref{beff_def1}) allows us to calculate an effective slip in
any direction given by an angle $\theta $, provided the two eigenvalues of the slip-length tensor, $b_{\mathrm{eff}%
}^{\parallel }$ ($\theta =0$) and $b_{\mathrm{eff}}^{\perp }$ ($\theta =\pi
/2$),  are known. The concept of an effective slip length tensor is general and can be applied
for an arbitrary channel thickness~\cite{harting.j:2012}, being a global characteristic
of a channel~\cite{vinogradova.oi:2011}, so that the eigenvalues
normally depend not only on the parameters of the heterogeneous
surfaces, but also on the channel thickness. However, for a thick
(compared to a texture period, $L$) channel we are interested in here they
become a characteristics of a heterogeneous interface solely.

In case of an anisotropic isolated surface (or a thick channel limit) with
a scalar local slip $b(y)$, varying in only one direction, the transverse component of the slip-length tensor was proven to be equal
to a half of the longitudinal one with twice larger local slip, $2b(y)$~\cite{asmolov:2012}
\begin{equation}
b_{\mathrm{eff}}^{\bot }\left[ b\left( y\right) /L\right] =\frac{b_{\mathrm{%
eff}}^{\parallel }\left[ 2b\left( y\right) /L\right]. }{2}  \label{aff}
\end{equation}%
A remarkable corollary of this relation is that the flow along any direction
of the one-dimensional surface can be easily determined, once the longitudinal component
of the effective slip tensor is found from the known spatially nonuniform
scalar slip.

One-dimensional superhydrophobic surfaces are very important for a class of phenomena, which involves
``transverse'' hydrodynamic couplings, where an applied pressure gradient or shear rate in
one direction generates flow in a different direction, with a
nonzero perpendicular component. This can be used to mix adjacent streams, control the
dispersion of plugs, and position streams within the cross section
of the channel~\cite{stroock2002b}. Such grooved surfaces can be easily prepared by modern lithographic methods~\cite{vinogradova.oi:2012}.

\begin{figure}
\begin{center}
\includegraphics [width=6.5 cm]{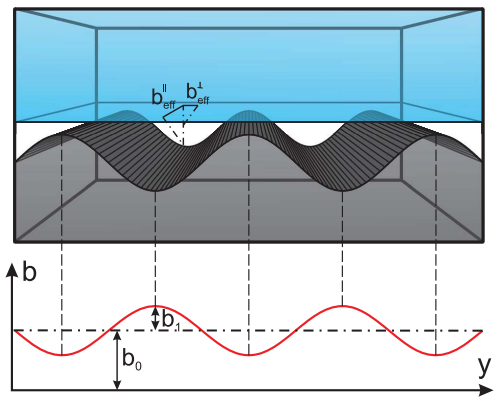}
   \end{center}
  \caption{(Color online) Sketch of the SH surface with a cosine relief, and its equivalent representation in terms of flow boundary conditions.  }
  \label{fig:geometry}
\end{figure}

Most of the prior work focussed on a flat, periodic, striped superhydrophobic surface, which corresponds to patterns of rectangular grooves. The
flow past such stripes was tackled theoretically ~\cite%
{lauga.e:2003,belyaev.av:2010a,feuillebois.f:2009,vinogradova.oi:2011,feuillebois.f:2010b,ng:2009}, and several numerical approaches have also been used either at the molecular scale, using molecular dynamics~\cite{priezjev.n:2011}, or at larger mesoscopic scales using finite element methods~\cite{priezjev.nv:2005,cottin.c:2004}, lattice Boltzmann~\cite{harting.j:2012}
and dissipative particle dynamics~\cite{zhou.j:2012} simulations. For a pattern composed
of no-slip ($b=0$) and perfect-slip ($b=\infty$) stripes, the
expression for the eigenvalues of the effective slip-length tensor
takes its maximum possible value and reads~\cite{philip.jr:1972,lauga.e:2003}
\begin{equation}
b_{\mathrm{ideal}}^{\parallel }=2b_{\mathrm{ideal}}^{\perp }=-\frac{L}{\pi }%
\ln \left[ \sec \left( \frac{\pi \left( 1-\phi \right) }{2}\right) \right] ,
\label{Phil72}
\end{equation}%
where $\phi $ is the fraction of the no-slip interface. In the limit of vanishing solid
fraction, it therefore predicts $b_{\mathrm{ideal}}^{\parallel}$ and $b_{\mathrm{ideal}}^{\perp }$ to depend only logarithmically on $\phi $ and scale as $-L\ln \phi $. At a qualitative level, this result means that the effective slip lengths essentially saturate at the value fixed by the period of the roughness. In case of stripes the perturbation of piecewise constant local slip has a step-like jump on the heterogeneity boundary, which leads to a singularity both in pressure and velocity gradient~\cite{asmolov:2012} by introducing an additional mechanism for a dissipation. It is natural to assume that an anisotropic one-dimensional texture with a continuous local slip could potentially lead to a larger effective tensorial slip.

In this paper we address the issue of the effective slip of flat surfaces with cosine variation in the local
slip length, which corresponds to modulated hydrophobic grooved
surfaces with a trapped gas layer (the Cassie state) as shown in Fig.~\ref{fig:geometry}. Flows over hydrophilic surfaces (the Wenzel state) with cosine surface relief of small amplitude have been studied by a number of authors~\cite{stroock2002b,hocking.lm:1976,wang2004,priezjev.nv:2006,niavarani.a:2010}. Previous studies of similar grooves in the Cassie state have investigated only small variations in local slip length~\cite{hocking.lm:1976,hendy2005effect}. We are unaware of any previous work that has studied the most interesting case of finite and large variations in the amplitude of a local cosine slip.

\section{Theory}

Consider a shear flow over a textured flat slipping plate, characterized by
a slip length $b(y)$, spatially varying in one direction, and the texture
varying over a period $L$ (as shown in Fig.~\ref{fig:geometry}). We use a rectangular coordinate system $(x,y,z)$
with origin at the wall. The $z-$ axis is perpendicular to the plate. Our
analysis is based on the limit of a thick channel or a single interface, so
that the velocity profile sufficiently far above the surface, at a height
large compared to $L$, may be considered as a linear shear flow. All
variables are non-dimensionalized using the texture period $L$ as the
characteristic length, the shear rate of the undisturbed flow $G$ and
the fluid viscosity $\mu .$

The dimensionless fluid velocity is sought in the form%
\begin{equation*}
\mathbf{v}=\mathbf{U}+\mathbf{u}_{\rm slip}+\mathbf{u}_{1}\left( x,y,z\right) ,
\end{equation*}%
where $\mathbf{U}=z\mathbf{e}_{l},\ l=x,y$ is the undisturbed linear shear
flow, and $\mathbf{e}_{l}\ $are the unit vectors parallel to the plate. The
perturbation of the flow, which is caused by the presence of the texture,
involves a constant slip velocity $\mathbf{u}_{\rm slip}=\left(
u_{\rm slip},v_{\rm slip},0\right) $ and a varying part $\mathbf{u}_{1}=\left(
u,v,w\right) $ of the velocity field. A periodic velocity $\mathbf{u}_{1}$
should decay at infinity and has zero average:%
\begin{equation}
\int_{0}^{1}\mathbf{u}_{1}dy=0.  \label{av_u}
\end{equation}%
At a small Reynolds number $Re=GL^{2}/\nu ,$ $\mathbf{u}_{1}$ satisfies
the dimensionless Stokes equations,%
\begin{gather}
\mathbf{\nabla }\cdot \mathbf{u}_{1}=0,  \label{Se} \\
\mathbf{\nabla }p-\Delta \mathbf{u}_{1}=\mathbf{0}.  \notag
\end{gather}

The boundary conditions at the wall and at infinity are defined in the usual
way as%
\begin{gather}
z=0:\quad \mathbf{u}_{\rm slip}+\mathbf{u}_{1\tau }-\beta \left( y\right) \frac{%
\partial \mathbf{u}_{1\tau }}{\partial z}=\beta \left( y\right) \mathbf{e}%
_{l},\   \label{bcu} \\
w=0,  \label{bcw0}
\end{gather}%
\begin{equation}
z\rightarrow \infty :\quad \mathbf{u}_{1}=\mathbf{0,}  \label{bci}
\end{equation}%
where $\mathbf{u}_{1\tau }=\left( u,v,0\right) $ is the velocity along the
wall and $\beta =b/L$ is the normalized local slip length. The eigenvalues
of the effective slip-length tensor can be obtained as the components of $%
\mathbf{u}_{\rm slip}:$
\begin{equation}
b_{\mathrm{eff}}^{\parallel }=Lu_{\rm slip},\quad b_{\mathrm{eff}}^{\perp
}=Lv_{\rm slip}.  \label{co_b}
\end{equation}

\section{Cosine slip length}

In this Section we consider a 1D periodic texture with the local slip length%
\begin{equation}  \label{eq:beta}
\beta =\beta _{0}+2\beta _{1}\cos \left( 2\pi y\right) .
\end{equation}%
The coefficients should satisfy $\beta _{0}\geq2\beta _{1}\geq0,~$in
order to obey $\beta \left( y\right) \geq0$ for any $y.$

The disturbance velocity field is presented in terms of Fourier series
as%
\begin{equation}
\mathbf{u}_{1}=\sum_{n=-\infty ,n\neq 0}^{\infty }\mathbf{u}^{\ast }\left(
n,z\right) \exp \left( i2\pi ny\right) ,\quad \mathbf{u}^{\ast }=\left(
u^{\ast },v^{\ast },w^{\ast }\right) .  \label{fu}
\end{equation}%
A general solution of the Stokes equations for the longitudinal flow, $%
\mathbf{U}=z\mathbf{e}_{x}$ decaying at infinity, reads \cite{asmolov:2012}%
\begin{equation}
u^{\ast }=X_{n}\exp \left( -2\pi \left\vert n\right\vert z\right) ,
\label{1x}
\end{equation}%
\begin{equation}
v^{\ast }=w^{\ast }=0,  \notag
\end{equation}%
and that for the transverse flow, $\mathbf{U}=z\mathbf{e}_{y}$ is
given by
\begin{equation*}
u^{\ast }=0,
\end{equation*}%
\begin{equation}
v^{\ast }=Y_{n}\exp \left( -2\pi \left\vert n\right\vert z\right) \left(
1-2\pi nz\right) ,  \label{1y}
\end{equation}%
\begin{equation}
w^{\ast }=-i2\pi nY_{n}z\exp \left( -2\pi \left\vert n\right\vert z\right) .
\label{1z}
\end{equation}%
The Fourier coefficients $X_{n}$ and $Y_{n}$ are determined from the Navier
slip boundary condition (\ref{bcu}).

\subsection{Longitudinal configuration}

Since the local slip length is an even function of $y,$ the solution (\ref%
{fu}) is also an even function. This requires $X_{n}=X_{-n},$ so it is
sufficient to evaluate $X_{n}$ for $n\geq 0.$ The Navier slip boundary
condition (\ref{bcu}) can be written, following to \cite{asmolov:2012}, in
terms of Fourier coefficients as a linear system
\begin{equation}
u_{\rm slip}=\beta _{0}-2e_{1}X_{1},  \label{ap1}
\end{equation}%
\begin{equation}
d_{1}X_{1}+e_{2}X_{2}=\beta _{1},  \label{ap2}
\end{equation}%
\begin{equation}
n>1:\quad e_{n-1}X_{n-1}+d_{n}X_{n}+e_{n+1}X_{n+1}=0,  \label{ap3}
\end{equation}%
\begin{equation}
d_{n}=1+2\pi n\beta _{0},\quad e_{n}=2\pi n\beta _{1}.  \label{a4}
\end{equation}%

A three-diagonal infinite linear system (\ref{ap1})-(\ref{ap3}) should be solved
numerically to find teh unknown $X_{n}$ by truncating the system.

 In the limit of large slip, $\beta _{0}>2\beta _{1}\gg 1,$ the asymptotic
solution to (\ref{ap1})-(\ref{ap3}) can also be constructed. To the leading
order in $\beta _{0}^{-1}$, the first term in (\ref{a4}) can be neglected
compared to the second one, and the system (\ref{ap2})-(\ref{ap3}) is
rewritten for new variables $t_{n}=2\pi nX_{n}$ as
\begin{equation}
t_{1}+\lambda t_{2}=\lambda ,  \label{ap4}
\end{equation}%
\begin{equation}
n>1:\quad \lambda t_{n-1}+t_{n}+\lambda t_{n+1}=0,  \label{ap5}
\end{equation}%
where $\lambda =\beta _{1}/\beta _{0}<1/2.$ The solution of the last system
is a geometric progression, $t_{n+1}=q^{n}t_{1},$ with%
\begin{equation*}
q=\frac{-\lambda ^{-1}+\sqrt{\lambda ^{-2}-4}}{2},
\end{equation*}%
\begin{equation*}
t_{1}=\frac{\lambda }{1+q\lambda }=\frac{\beta _{1}}{\beta _{0}+q\beta _{1}}.
\end{equation*}%
Therefore, the final expression for the slip length, $b_{\mathrm{eff}%
}^{\parallel }=Lu_{\rm slip}$, in view of (\ref{ap1}), takes the following form at $\beta
_{0}>2\beta _{1}\gg 1:$%
\begin{equation}
b_{\mathrm{eff}}^{\parallel }=\sqrt{b_{0}^{2}-4b_{1}^{2}}.  \label{be_par}
\end{equation}

The asymptotic solution for the velocity field can also be derived from Eqs.(%
\ref{fu}) and (\ref{1x}). The normal component of velocity gradient can be
written as%
\begin{equation}
\frac{\partial u}{\partial z}=-2t_{1}\sum_{n=1}^{\infty }q^{n-1}\mathrm{Re}%
\left\{ \exp \left[ 2\pi n\left( iy-z\right) \right] \right\} .  \label{dudz}
\end{equation}%
The factor $2$ is due to the contribution of $n<0.$ The sum in (\ref{dudz})
is also a geometric progression, so that%
\begin{equation}
\frac{\partial u}{\partial z}=-\frac{2t_{1}\exp \left( -2\pi z\right) \left[
\cos \left( 2\pi y\right) -q\exp \left( -2\pi z\right) \right] }{s},
\label{dul}
\end{equation}%
\begin{equation*}
s=1-2q\cos \left( 2\pi y\right) \exp \left( -2\pi z\right) +q^{2}\exp \left(
-4\pi z\right) .
\end{equation*}%
The last equation can be integrated over $z$ to give%
\begin{equation}
u=-\frac{t_{1}}{2\pi q}\ln s.  \label{ul}
\end{equation}%
The asymptotic solutions (\ref{be_par}), (\ref{dul}) and (\ref{ul}) predict
well the numerical data even at finite $b_{0}$ (see next Section).

\subsection{Transverse configuration}

It was found by \cite{asmolov:2012} that the velocity components for the
transverse flow can be expressed in terms of the longitudinal one calculated
for twice larger local slip, $u_{2}=u\left[ 2\beta \left( y\right) \right] :$%
\begin{equation}
v_{\rm slip}=\frac{u_{slip,2}}{2},  \label{v_slip}
\end{equation}

\begin{eqnarray}
v &=&\frac{1}{2}\left( u_{2}+z\frac{\partial u_{2}}{\partial z}\right) ,
\label{double} \\
w &=&-\frac{z}{2}\frac{\partial u_{2}}{\partial y}.
\end{eqnarray}
Using (\ref{co_b}) and (\ref{v_slip}) we derive%
\begin{equation}
b_{\mathrm{eff}}^{\perp }=b_{\mathrm{eff}}^{\parallel }=\sqrt{%
b_{0}^{2}-4b_{1}^{2}}.  \label{be_per}
\end{equation}
Thus, the texture is isotropic at large $b_{0}$. One can demonstrate that the conclusion about isotropy of the slip-length tensor is general and valid for any textures (see Appendix~\ref{A1}).

The values $\lambda ,q,t_{1}$ remain the same for the double slip length
since they depend on the ratio $\beta _{1}/\beta _{0}$ only. As a result, we
obtain%
\begin{equation*}
u_{2}=u=-\frac{t_{1}}{2\pi }\ln s,
\end{equation*}%
\begin{eqnarray}
v &=&-\frac{t_{1}}{4\pi q}\ln s \\
&-&\frac{zt_{1}\exp \left( -2\pi z\right) \left[ \cos \left( 2\pi y\right)
-q\exp \left( -2\pi z\right) \right] }{s},  \notag \\
w &=&\frac{zt_{1}\exp \left( -2\pi z\right) \sin \left( 2\pi y\right) }{s}.
\label{v}
\end{eqnarray}

\section{Simulation method}

\label{sec:lbm}

For the modelling of fluid flow in a system of two parallel
plates, we employ the lattice Boltzmann (LB) method ~\cite{bib:succi-01}.
Lattice Boltzmann methods are derived by a phase space discretization of the
kinetic Boltzmann equation
\begin{equation}
\left[ \frac{\partial }{\partial t}+\mathbf{v}\cdot \nabla _{\mathbf{r}}%
\right] f(\mathbf{r,v},t)=\mathbf{\Omega },  \label{eq:boltzmann}
\end{equation}%
which expresses the dynamics of the single particle probability density $f(%
\mathbf{r},\mathbf{v},t)$. Therein, $\mathbf{r}$ is the position, $\mathbf{v}
$ the velocity, and $t$ the time. The left-hand side models the propagation
of particles in phase space, the collision operator $\mathbf{\Omega }$ on
the right hand side accounts for particle interactions.

Constructing the lattice Boltzmann equation, the time $t$, the position $%
\mathbf{r}$, and the velocity $\mathbf{v}$ are discretized. This discrete
variant of Eq.~(\ref{eq:boltzmann})
\begin{equation}
\begin{array}{cc}
f_{k}(\mathbf{r}+\mathbf{c}_{k},t+1)-f_{k}(\mathbf{r},t)=\Omega _{k}, &
k=0,1,\dots ,B,%
\end{array}%
\end{equation}%
describes the kinetics in discrete time- ($\Delta t$) and space-units ($%
\Delta x$). We employ a widely used three-dimensional lattice with $B=18$
discrete non-zero velocities (D3Q19) which is chosen to carry sufficient symmetry to
allow for a second order accurate solution of the Navier-Stokes equations.
Here, for $\Omega $, we choose the Bhatnagar-Gross-Krook (BGK) collision
operator~\cite{bib:bgk}
\begin{equation}
\Omega _{k}=-\frac{1}{\tau }\left( f_{k}(\mathbf{r},t)-f_{k}^{eq}(\mathbf{v}(%
\mathbf{r},t),\rho (\mathbf{r},t))\right) \mbox{ ,}  \label{Omega}
\end{equation}%
which assumes relaxation on a linear timescale $\tau $ towards a discretized
local Maxwell-Boltzmann distribution $f_{k}^{eq}$. The kinematic viscosity $%
\nu =\frac{2\tau -1}{6}$ of the fluid is related to the relaxation time
scale. In this study it is kept constant at $\tau =1.0$.

Stochastic moments of $f$ can be related to physical properties of the
modelled fluid. Here, conserved quantities, like the fluid density $\rho (%
\mathbf{r},t)=\rho _{0}\sum_{k}f_{k}(\mathbf{r},t)$ and momentum $\rho (%
\mathbf{r},t)\mathbf{u}(\mathbf{r},t)=\rho _{0}\sum_{k}c_{k}f_{k}(\mathbf{r}%
,t)$, with $\rho _{0}$ being a reference density, are of special interest.

Slip over hydrophobic surfaces is commonly modelled by introduction of a
phenomenological repulsive force ~\cite%
{bib:jens-kunert-herrmann:2005,bib:jens-jari:2008,bib:zhu-tretheway-petzold-meinhart-2005,bib:benzi-etal-06,bib:zhang-kwok-04}%
. The magnitude of interactions between different components and surfaces,
as determined by simulation parameters allows to specify arbitrary
contact-angles ~\cite%
{benzi-etal-06b,bib:huang-thorne-schaap-sukop-2007,bib:jens-schmieschek:2010,bib:jens-kunert-herrmann:2005}%
. Other approaches include boundary conditions taking into account specular
reflections~\cite{succi02,bib:tang-tao-he-2005,bib:sbragaglia-succi-2005},
or diffuse scattering~\cite%
{bib:ansumaili-karlin-2002,bib:sofonea-sekerka-2005,bib:niu-shu-chew-2004},
respectively. The strategy applied for this work employs a second order
accurate on-site fixed velocity boundary condition to simulate wall
slippage. Here, the velocity at the boundary is set proportional to the
local stress imposed by the flow field as well as a slip length parameter
adjusting the stress-response. For the details of the implementation we
refer the reader to~\cite{HechtHarting2010,ahmed.nk:2009}. Local slip
lengths are calculated according to Eq.~(\ref{eq:beta}). Varying slip
patterns are applied to the $x$-$y$-plane at $z=0$. Periodic boundary
conditions are employed in $x$ and $y$-direction, thus reducing the
simulation domain to a pseudo-2D system. By exploiting the periodic
boundaries, only one single period needs to be resolved. The Couette flow is driven
by applying a constant velocity of $v=0.1$ (in lattice units) in  the
$x$-$y$-plane at $z=z_{\rm max}$.

% Optional creating a new image depicting the dataset to be fitted including
% the Cosine hull curve of the flow field

The resolution of the simulated system is given by the lattice constant
\begin{equation}
\Delta x = \frac{H}{\mathcal{N}},
\end{equation}
where $\mathcal{N}$ is the number of discretization points used to resolve
the height of the channel. As we consider thick channels,
  we choose a height to periodicity ratio of $H/L=10$.

The number of timesteps required to reach a steady state depends on the
channel height, the velocity of the flow as determined by the driving
acceleration as well as the fraction of slip and no slip area at the
surface. We find that in order to reduce the deviation from theoretical
predicted values below 5 percent, a domain size of $96\times 1\times
960\Delta x^{3}$ is required. For this geometrical setup, with a shear rate in the order of $\dot{\gamma} = 1\cdot
10^{-4}$  (in lattice units), a simulation time in the
order of 10 million timesteps is needed to reach the steady state. We compare
the theoretical predictions by measurements of $b_{\mathrm{eff}}$ obtained from
velocity profiles. The profile of the linear shear velocity is fit by a linear
function in the region far from the surface pattern. From this fit the
effective slip lengths and shear rate are determined.

%To vary the inclination of the pattern orientation relative to the flow, the
%simulated angle of influx is rotated in the $x-y$-plane rather than the
%pattern orientation itself. This minimizes discretization errors due to the
%underlying regular lattice occurring in case of a rotated surface pattern.
%This technique has to be accounted for regarding data extracted from the
%lattice by subsequent projection of the measured values onto the flow
%direction.

\section{Results and discussion}

In this section, we present the LB simulation results and
compare them with predictions of the continuous theory.

\begin{figure}[tbp]
\begin{center}
\includegraphics[scale=0.45]{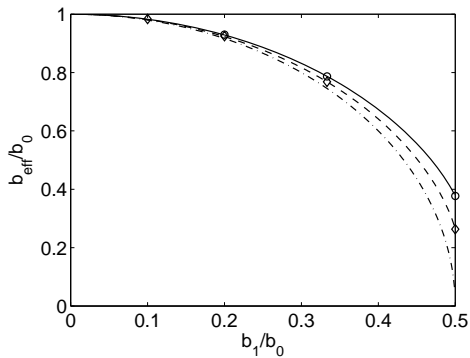}
\end{center}
\caption{Eigenvalues of the effective slip-length tensor as a function
  of $b_1$ simulated at fixed $b_{0}=1$ (symbols). The longitudinal effective slip length, $b_{\mathrm{eff}%
}^{\parallel }$, is shown by circles, and the transverse effective
slip, $b_{\mathrm{eff}}^{\perp }$ is presented by diamonds. Solid and
dashed curves denote the corresponding theoretical values obtained by
numerical Fourier-series solutions. The asymptotic (isotropic)
solution, Eq. (\protect\ref{be_per}), expected in the limit $b_{0}\gg
L$ is shown by the dash-dotted line.}
\label{be}
\end{figure}

We start with varying the amplitude of cosine perturbations of the slip length, $b_{1}$, at fixed $b_{0}=1$.  Fig. \ref{be} shows simulation data for $b_{\mathrm{eff}%
}^{\parallel }\ $ and $b_{\mathrm{eff}}^{\perp }$ as a function
of $b_{1}/b_0$.  These results show that the largest possible value of
$b_{\mathrm{eff}}/b_0$ is attained when $b_1=0$, i.e. for a smooth hydrophobic surface with $b(y)=b_0$. In this situation the effective slip is (obviously) isotropic and equal to the area-averaged slip $b_0$. When increasing the amplitude $b_1$, there is a small anisotropy of the flow, and the eigenvalues of the slip-length tensor decrease. Therefore, in the presence of a cosine variation in slip length the effective slip always becomes smaller than average. This conclusion is consistent with earlier observations made for different textures~\cite{alexeyev:96,vinogradova.oi:2011}. To obtain theoretical values, the linear system, Eqs.~(\ref{ap1})-(\ref{ap3}) has been solved
numerically by using the IMSL-DLSLTR routine. We see that the agreement is excellent for all $b_1/b_0$,
indicating that our asymptotic theory is extremely accurate, and
confirming the relation (\ref{v_slip}) between the longitudinal and
transverse slip lengths. Also included in Fig.~\ref{be} is the
asymptotic formula (\ref{be_per}) obtained in the limit of large
$b_0$. Note that this formula is surprisingly accurate even in the case of finite $b_{0},$ except for the
texture with $b_{1}/b_{0}=1/2$ (no-slip point at $y=1/2$).

\begin{figure}[tbp]
\begin{center}
\includegraphics[scale=0.45]{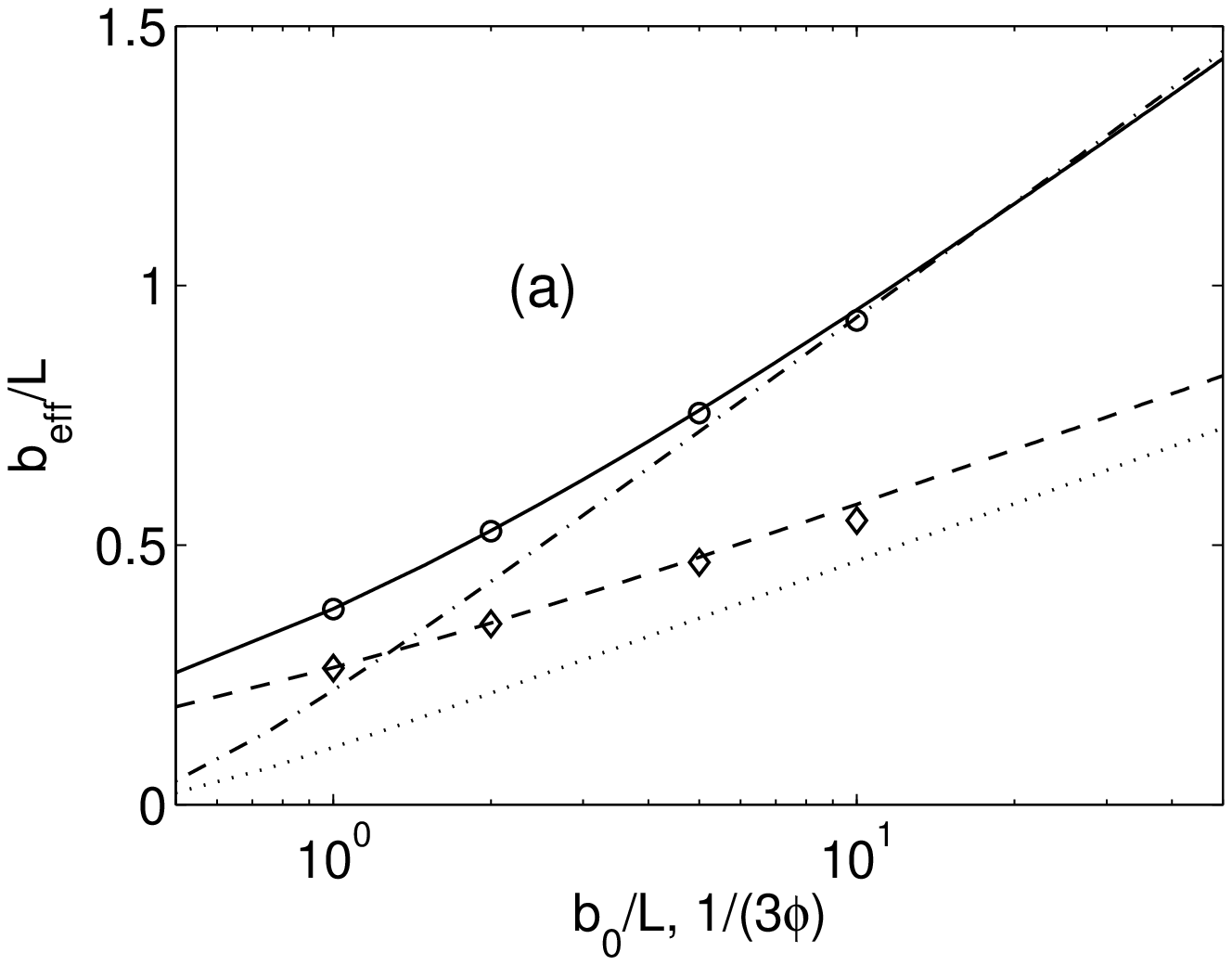}
\includegraphics[scale=0.45]{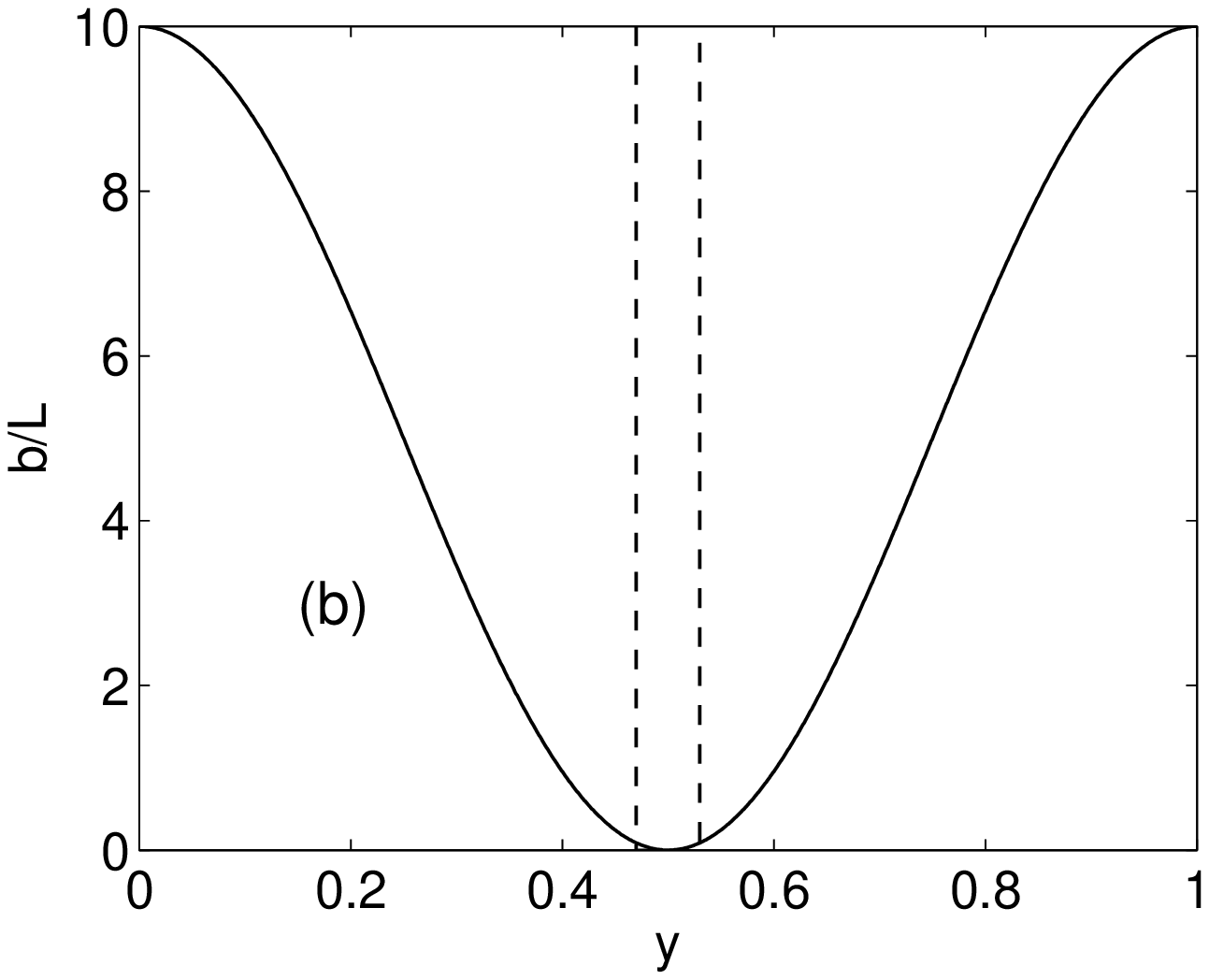}
\end{center}
\caption{$\left( a\right) $ Effective slip lengths computed for $b_{1}/b_{0}=0.5$, which correspond to a texture with no-slip lines, vs. average slip. The notations are the same as in Fig. \protect\ref{be}.
Dash-dotted and dotted lines show the effective lengths for longitudinal and
transverse stripes as a function of $1/(3 \phi)$ calculated with Eq.(\protect\ref{Phil72}).
$\left( b\right) $ The cosine
profile of the local slip length with $b_{0}/L=5, \ b_{1}/b_{0}=0.5$
(solid curve) and the stripe profile with $\protect\phi =0.06 $ (dashed
line) with the same longitudinal effective slip lengths.}
\label{noslip}
\end{figure}

Fig.~\ref{noslip} $\left( a\right) $ shows the simulation data for
effective slip lengths as a function of average slip, $b_0/L$, for a
texture with the no-slip point $\left( b_{1}/b_{0}=1/2\right)$. Also
included are theoretical (Fourier series) curves. The fits are quite good for  $b_{0}/L$ up to 10, but at larger average slip there is
some discrepancy. The simulation results for $b_{\mathrm{eff}}^{\parallel }$ and $b_{\mathrm{eff}}^{\perp }$ give smaller values than predicted by the theory. A possible explanation for this discrepancy is that the major contribution to the shear stress at large $%
b_{0}/L$ and $b_{1}/b_{0}=1/2$ comes from a very small region near the no-slip
point (as we discuss below). The discretization error of
the LB simulation becomes maximal in this region, and is particularly
pronounced for the velocity gradients of systems with large effective
slip. While we observe deviations around the no slip extremal value,
the curves converge fast when stepping away from it and the excellent
agreement of the measured effective slip suggest that the influenece
of discretization errors on the mean flow is negligible at the
resolution used. The asymptotic formula, Eq.(\ref{be_per}), predicts $b_{\mathrm{eff}}=0$.
This likely indicates that in this situation it is necessary to construct the second-order term of expansions for eigenvalues. At relatively large $b_{0}/L$, the
effective slip lengths can be well fitted as%
\begin{eqnarray}
b_{\mathrm{eff}}^{\parallel }/L &\simeq&0.1871+0.3175\ln \left(
b_{0}/L+1.166\right) ,  \label{fit} \\
b_{\mathrm{eff}}^{\perp }/L &\simeq&0.2036+0.158\,8\ln \left(
b_{0}/L+0.583\right) .
\notag
\end{eqnarray}
In other words, they scale as $\ln (b_{0}/L)$ at large $b_{0}/L.$

Since the effective slip lengths for a texture decorated with
perfect-slip stripes,  Eq.(\ref{Phil72}), also shows a logarithmic
growth (with $\phi $), in order to compare these two one-dimensional
anisotropic textures the theoretical curve for stripes is included in
Fig.~\ref{noslip}$\left(a\right)$. It can be seen that in the limit
of large average slip the asymptotic curves for longitudinal effective slip for stripes and cosine texture nearly coincide. This means that both textures generate the same forward flow in the longitudinal direction. Simple estimates suggest $%
b_{\mathrm{eff}}^{\parallel }\left( \beta _{0}\right) \simeq b_{\mathrm{ideal%
}}^{\parallel }\left[ 1/\left( 3\beta _{0}\right) \right].$ Perhaps the most interesting and important aspect of
this observation is that, from the point of view of the longitudinal effective slip, the ``wide'' cosine texture with $\beta
_{0}=5$ taken for our numerical example is equivalent to the patterns of stripes with the extremely low fraction of no-slip regions, $\phi =0.06$  (see Fig.~\ref{noslip}$\left( b\right) $). These results may guide the design of superhydrophobic surfaces for large forward flows in microfluidic devices. Note, however, that in the situation when longitudinal slip for both textures are similar, the cosine texture shows a larger transverse effective slip as seen in Fig.~\ref{noslip}$\left( a\right) $. This means that textures with the cosine
variation in the local slip length are less anisotropic than stripes, and $b_{\mathrm{eff}}^{\parallel
}/b_{\mathrm{eff}}^{\perp }<b_{\mathrm{ideal}}^{\parallel }/b_{\mathrm{ideal}%
}^{\perp }=2$. Therefore, cosine textures are less optimal for a generation of robust transverse flows as compared with a sharp-edge stripe geometry.

\begin{figure}[tbp]
\includegraphics[scale=0.45]{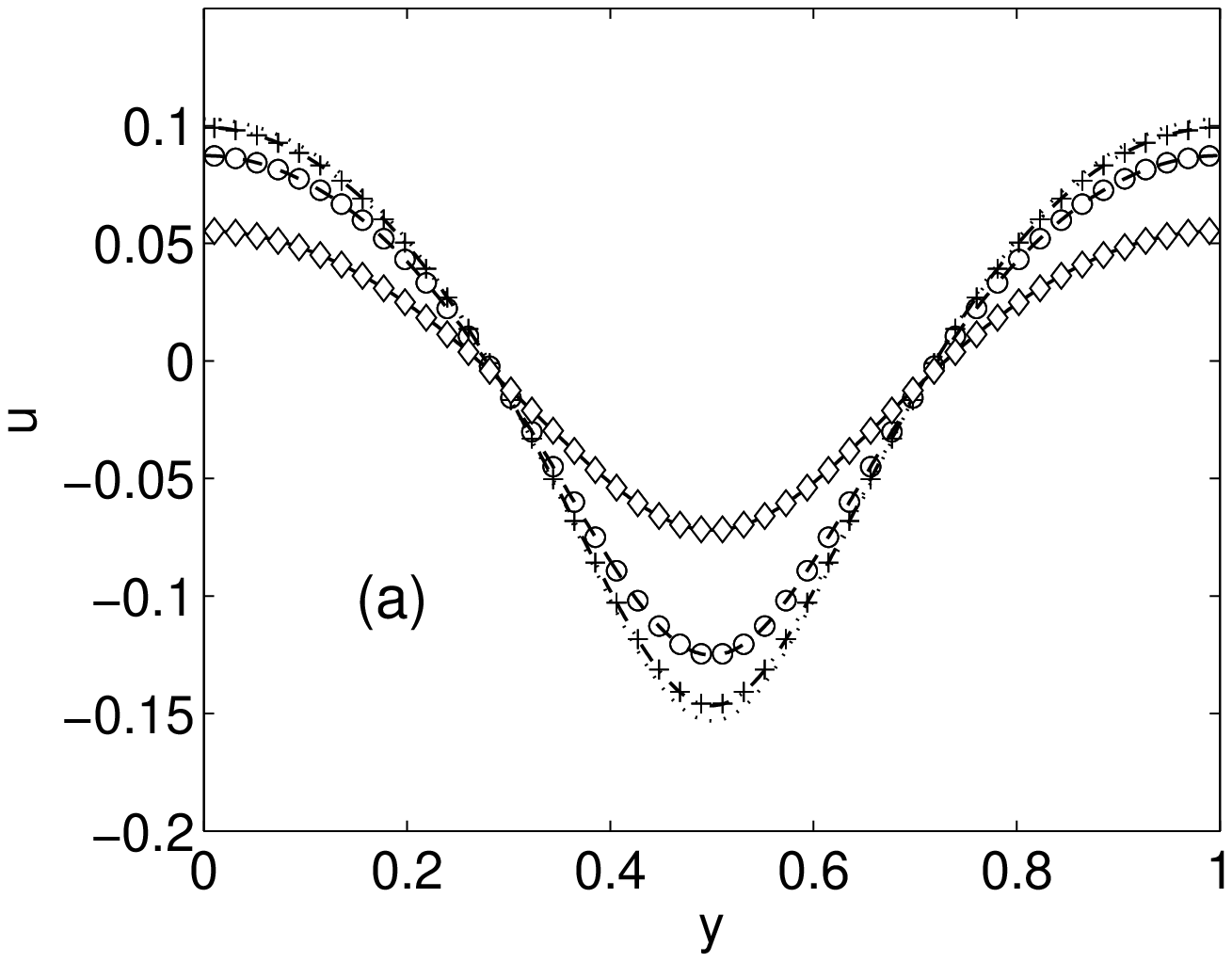}
\includegraphics[scale=0.45]{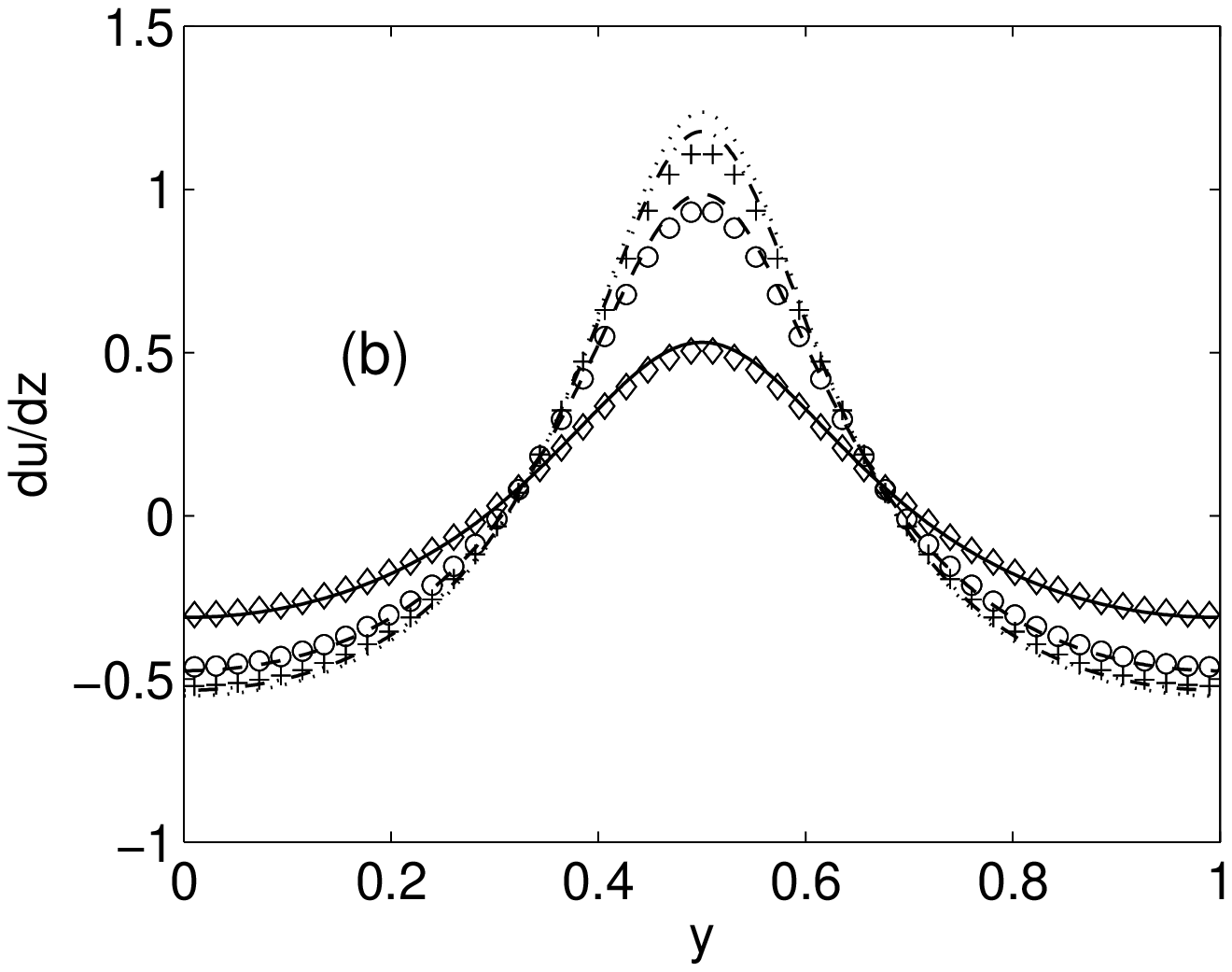}
\caption{$\left( a\right) $ The velocities and $\left( b\right) \ $the
normal velocity gradients along the wall for the textures with $%
b_{1}/b_{0}=1/3,\ b_{0}/L=0.2$ (solid curve, diamonds)$,\ b_{0}/L=1$ (dashed curve, circles)$,\ b_{0}/L=5$
(dash-dotted line, crosses). Dotted curves show predictions of asymptotic formulae, Eqs.~(\protect\ref{ul}) and (\protect\ref%
{dul}).}
\label{u3}
\end{figure}

\begin{figure}[tbp]
\begin{center}
\includegraphics[scale=0.45]{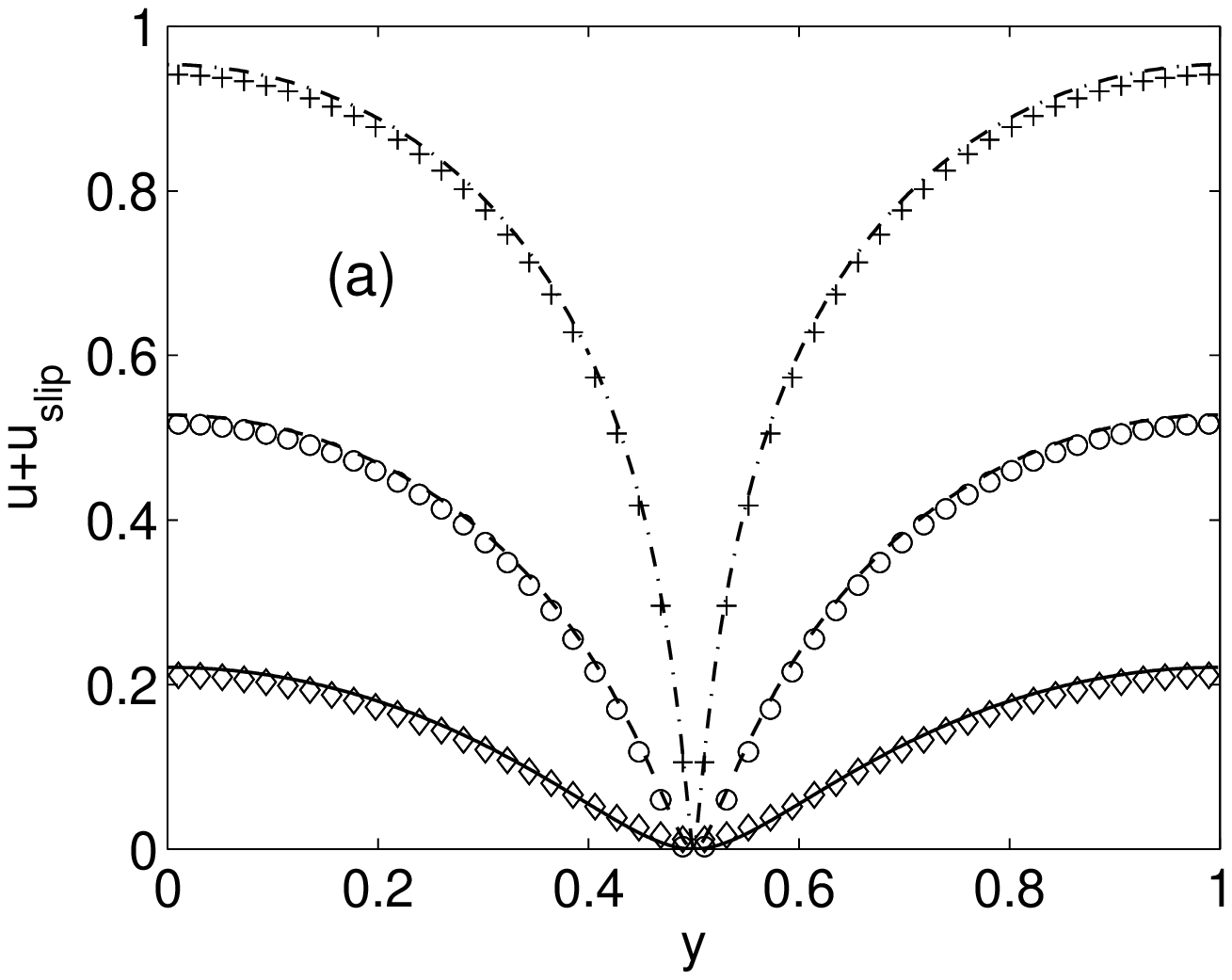} %
\includegraphics[scale=0.45]{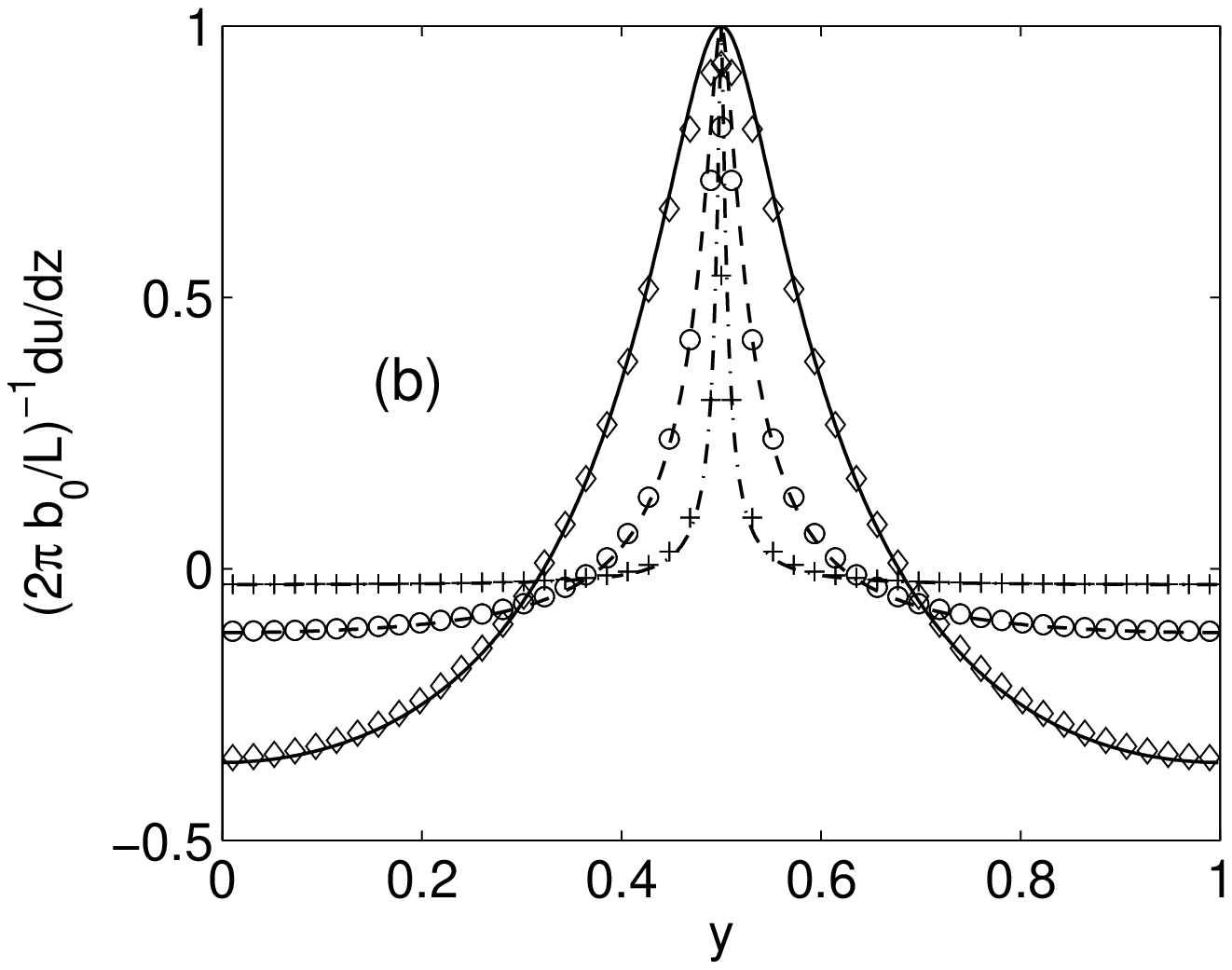}
\end{center}
\caption{$\left( a\right) $ The velocities and $\left( b\right) \ $the
normal velocity gradients along the wall for the no-slip textures ($%
b_{1}/b_{0}=1/2$). Other notations are the same as in Fig.~\protect\ref{u3}}
\label{u2}
\end{figure}

\begin{figure}[tbp]
\begin{center}
\includegraphics[scale=0.45]{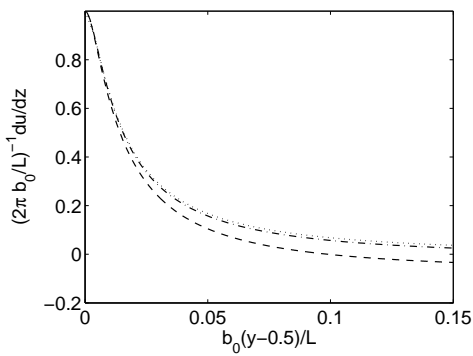}
\end{center}
\caption{The normal velocity gradients for the no-slip textures with large amplitudes of slip length variation, $b_{0}/L=1$ (dashed curve)$,\ b_{0}/L=5$
(dash-dotted curve),$\ b_{0}/L=20$ (dotted curve) as functions of stretched coordinates.}
\label{duc}
\end{figure}

The flow direction is associated with hydrodynamic
pressures in the film, which is related to the
heterogeneous slippage at the wall. Fig.~\ref{u3}  shows the profiles of the velocity and of the normal
velocity gradient along the wall for different $\beta _{0}$ and $\beta
_{1}/\beta _{0}>1/2$. The velocity dependence $u\left( x,y,0\right) $ is smooth,
and $\frac{\partial u}{\partial z}\left( x,y,0\right) $ is finite for any $%
\beta _{0}$ and $\beta _{1}$, unlike the striped textures with piecewise-constant $%
\beta $~\cite{asmolov:2012}. Asymptotic predictions (\ref{ul}) and (\ref%
{dul}) are in a good agreement with numerical
results and simulation data.

Similar theoretical and simulation results, but obtained for a texture with no-slip point, $\beta
_{1}/\beta _{0}=1/2$, are shown in Fig.~\ref{u2}. In this situation we find that $\partial
u/\partial z\left( 1/2\right) =2\pi \beta_{0}$ for all $\beta_{0}$.

Finally, we should like to stress that a very small region near the no-slip point gives a main contribution to the shear stress at large $b_{0}/L.$ For the major portion of the texture far from
this region, we have $\partial u/\partial z\simeq -1,$ so that the total
shear stress is zero, and this part of the texture is shear-free. Since the
maximum values of the normal velocity gradients grow like $b_{0}/L$ one can
expect that a length scale of this small region is $%
L^{2}/b_{0}\ll L,$ or, equivalently, the curvature radius, $r=\left(
d^{2}b/dy^{2}\right) ^{-1}=L^{2}/\left( 4\pi ^{2}b_{0}\right) \ll L,$ at the
no-slip point. The validity of this assumption is justified in Fig. \ref{duc}, where the
gradients for several values of $b_{0}/L$ versus stretched coordinates, $%
b_{0}\left( y-0.5\right) /L,$ are presented. The curves are very close
for $b_{0}/L\geq5$. Therefore, the normal gradient distribution and the
dimensionless effective slip length in this case are controlled by the ratio $r/L$ only. These conclusions can be extended to any $b\left( y\right) $ characterized by a small radius $r=\left(d^{2}b/dy^{2}\right) ^{-1}\ll L$ near the no-slip point and by a large slope, of the order of $L/r\gg 1$ or larger, far
from it.

\iffalse%%%%%%%%%%%%%%%%%%%%%%%%%%%%%%%%%%%%%%%%%%%%%%%%%%%%

Eqs. (\ref{fit}) then can be written in terms of the radius $r$ as%
\begin{eqnarray}
b_{\mathrm{eff}}^{\parallel }/L &=&-0.98+0.3175\ln \left( L/r+46.03\right) ,
\label{fir} \\
b_{\mathrm{eff}}^{\perp }/L &=&-0.38+0.158\,8\ln \left( L/r+23.01\right) .
\notag
\end{eqnarray}%
The above conclusion can be extended to
any dependence $b\left( y\right) $ characterizing by a small radius $r=\left(d^{2}b/dy^{2}\right) ^{-1}\ll L$
near the no-slip point and by a large slope, of the order of $L/r\gg 1$ or larger, far
from it. Eqs. (\ref{fir}) can be also applied to estimate $\mathbf{b}_{%
\mathrm{eff}}$ for such textures.

\fi%%%%%%%%%%%%%%%%%%%%%%%%%%%%%%%%%%%%%%%%%%%%%

\section{Conclusion}
We have investigated shear flow past a super-hydrophobic surface with a
cosine variation of the local slip length, and have evaluated resulting effective
slippage and the flow velocity. We have found that the cosine texture can
provide a very large effective (forward) slip, but generates a smaller transverse
velocity to the main (forward) flow than discrete stripes considered earlier. Our approximate
formulae for longitudinal and transversal directional effective slip lengths
are validated by means of lattice Boltzmann simulations. Excellent quantitative
agreement is found for the effective slippage as well as for the flow field.
Slight deviations of the observed velocity gradient close to the no-slip
extremal value can be explained by discretization errors.

%\section*{Acknowledgements}
\begin{acknowledgments}
This research was partly supported by the Russian Academy of Science (RAS)
through its priority program `Assembly and Investigation of Macromolecular
Structures of New Generations', by the Netherlands Organization for Scientific
Research (NWO/STW VIDI), and by the German Science Foundation (DFG) through its
priority program `Micro- and nanofluidics'. We acknowledge computing resources
from the J\"ulich Supercomputing Center and the Scientific Supercomputing
Center Karlsruhe.
\end{acknowledgments}

\appendix
\section{Effective slip for a flow past strongly slipping anisotropic patterns}\label{A1}

In this Appendix we give some simple arguments showing that the flow becomes isotropic in case of large local slip,  $%
\beta _{\min }=\min \left[ \beta \left( y\right) \right] \gg 1.$ In this
situation for arbitrary
local slip length $\beta \left( y\right)$ the
leading-order $\mathbf{b}_{\mathrm{eff}}$ can be
obtained directly from the local slip boundary condition, Eq.(\ref{bcu}), without a need to solve the Stokes equations.  Indeed, if $\beta \gg 1$,  Eq. (\ref{bcu}) requires  that the velocity at the surface is large,
\begin{equation*}
\left\vert \mathbf{u}_{\rm slip}+\mathbf{u}_{1\tau }\left( x,y,0\right)
\right\vert \sim \beta _{\min }\gg 1,
\end{equation*}%
but the normal gradient is finite, $\left\vert \partial \mathbf{u}_{1\tau }\left( x,y,0\right) /\partial
z\right\vert \sim 1.$
This can be fulfilled if only $\left\vert \mathbf{u}_{\rm slip}\right\vert \sim \beta _{\min }\gg 1,\quad
\left\vert \mathbf{u}_{1\tau }\right\vert \sim 1.$
One can then divide Eq. (\ref{bcu}) by $\beta $ to obtain%
\begin{equation*}
z=0:\quad \frac{\mathbf{u}_{\rm slip}}{\beta \left( y\right) }-\frac{\partial
\mathbf{u}_{1\tau }}{\partial z}-\mathbf{e}_{l}=O\left( \beta _{\min
}^{-1}\right) .
\end{equation*}%
Finally, by averaging of the last equation over the texture periods, and
by using Eqs.(\ref{av_u}) and Eq.(\ref{co_b}) we can easily derive an expression for a leading-order effective slip:%
\begin{equation}
b_{\mathrm{eff}}^{\perp }\simeq b_{\mathrm{eff}}^{\parallel }\simeq L\left[ \int_{0}^{1}%
\frac{dy}{\beta \left( y\right) }\right] ^{-1}.  \label{u_sl}
\end{equation}
It can be verified that our result, Eq.(\ref{be_per}), is consistent with %
(\ref{u_sl}). Note that the expression
\begin{equation}
b_{\mathrm{eff}}^{\parallel }\simeq b_{\mathrm{eff}}^{\bot }\simeq \left(
\frac{\phi }{b_{0}}+\frac{1-\phi }{b_{1}}\right) ^{-1},  \label{Ng}
\end{equation}%
which has been obtained for stripes with $%
b_{0},b_{1}\gg L$~\cite{cottin.c:2004} and is similar to the addition rule for resistors in parallel,
 also satisfies Eq.(\ref{u_sl}).

\bibliographystyle{rsc}
\bibliography{asmolov}

\providecommand{\url}[1]{\texttt{#1}}
\begin{thebibliography}{10}

\bibitem{quere.d:2008}
D.~Quere, \emph{Annu. Rev. Mater. Res.}, 2008, \textbf{38}, 71--99.

\bibitem{bocquet2007}
L.~{Bocquet} and J.~L. Barrat, \emph{Soft Matter}, 2007, \textbf{3}, 685--693.

\bibitem{vinogradova.oi:2011}
O.~I. Vinogradova and A.~V. Belyaev, \emph{J. Phys.: Condens. Matter}, 2011,
  \textbf{23}, 184104.

\bibitem{rothstein.jp:2010}
J.~P. Rothstein, \emph{Annu. Rev. Fluid Mech.}, 2010, \textbf{42}, 89--109.

\bibitem{bhushan.b:2011}
B.~Bhushan, \emph{Beilstein J. Nanotechnol.}, 2011, \textbf{2}, 66--84.

\bibitem{vinogradova.oi:2012}
O.~I. Vinogradova and A.~L. Dubov, \emph{Mendeleev Commun.}, 2012, \textbf{22},
  229--236.

\bibitem{vinogradova.oi:1999}
O.~I. Vinogradova, \emph{Int. J. Miner. Proc.}, 1999, \textbf{56}, 31--60.

\bibitem{lauga2005}
E.~Lauga, M.~P. Brenner and H.~A. Stone, in \emph{Handbook of Experimental
  Fluid Dynamics}, ed. C.~Tropea, A.~Yarin and J.~F. Foss, Springer, NY, 2007,
  ch.~19, pp. 1219--1240.

\bibitem{vinogradova.oi:1995a}
O.~I. Vinogradova, \emph{Langmuir}, 1995, \textbf{11}, 2213.

\bibitem{andrienko.d:2003}
D.~Andrienko, B.~D\"unweg and O.~I. Vinogradova, \emph{J. Chem. Phys.}, 2003,
  \textbf{119}, 13106.

\bibitem{vinogradova.oi:2003}
O.~I. {Vinogradova} and G.~E. {Yakubov}, \emph{Langmuir}, 2003, \textbf{19},
  1227--1234.

\bibitem{vinogradova.oi:2009}
O.~I. Vinogradova, K.~Koynov, A.~Best and F.~Feuillebois, \emph{Phys. Rev.
  Lett.}, 2009, \textbf{102}, 118302.

\bibitem{charlaix.e:2005}
C.~Cottin-Bizonne, B.~Cross, A.~Steinberger and E.~Charlaix, \emph{Phys. Rev.
  Lett.}, 2005, \textbf{94}, 056102.

\bibitem{joly.l:2006}
L.~Joly, C.~Ybert and L.~Bocquet, \emph{Phys. Rev. Lett.}, 2006, \textbf{96},
  046101.

\bibitem{choi.ch:2006}
C.~H. Choi, U.~Ulmanella, J.~Kim, C.~M. Ho and C.~J. Kim, \emph{Phys. Fluids},
  2006, \textbf{18}, 087105.

\bibitem{joseph.p:2006}
P.~Joseph, C.~Cottin-Bizonne, J.~M. Beno\v{\i}, C.~Ybert, C.~Journet,
  P.~Tabeling and L.~Bocquet, \emph{Phys. Rev. Lett.}, 2006, \textbf{97},
  156104.

\bibitem{Kamrin_etal:2010}
K.~Kamrin, M.~Z. Bazant and H.~A. Stone, \emph{J.~Fluid Mech.}, 2010,
  \textbf{658}, 409--437.

\bibitem{Bazant08}
M.~Z. Bazant and O.~I. Vinogradova, \emph{J.~Fluid Mech.}, 2008, \textbf{613},
  125--134.

\bibitem{harting.j:2012}
S.~Schmieschek, A.~V. Belyaev, J.~Harting and O.~I. Vinogradova, \emph{Phys.
  Rev. E}, 2012, \textbf{85}, 016324.

\bibitem{asmolov:2012}
E.~S. Asmolov and O.~I. Vinogradova, \emph{J. Fluid Mech.}, 2012, \textbf{706},
  108--­117.

\bibitem{stroock2002b}
A.~D. {Stroock}, S.~K. {Dertinger}, G.~M. {Whitesides} and A.~{Ajdari},
  \emph{Anal. Chem.}, 2002, \textbf{74}, 5306--5312.

\bibitem{lauga.e:2003}
E.~Lauga and H.~A. Stone, \emph{J.~Fluid Mech.}, 2003, \textbf{489}, 55--77.

\bibitem{belyaev.av:2010a}
A.~V. Belyaev and O.~I. Vinogradova, \emph{J.~Fluid Mech.}, 2010, \textbf{652},
  489--499.

\bibitem{feuillebois.f:2009}
F.~Feuillebois, M.~Z. Bazant and O.~I. Vinogradova, \emph{Phys. Rev. Lett.},
  2009, \textbf{102}, 026001.

\bibitem{feuillebois.f:2010b}
F.~Feuillebois, M.~Z. Bazant and O.~I. Vinogradova, \emph{Phys. Rev. E}, 2010,
  \textbf{82}, 055301(R).

\bibitem{ng:2009}
C.~Ng and C.~Wang, \emph{Phys. Fluids}, 2009, \textbf{21}, 013602.

\bibitem{priezjev.n:2011}
N.~V. Priezjev, \emph{J. Chem. Phys.}, 2011, \textbf{135}, 204704.

\bibitem{priezjev.nv:2005}
N.~V. Priezjev, A.~A. Darhuber and S.~M. Troian, \emph{Phys. Rev. E}, 2005,
  \textbf{71}, 041608.

\bibitem{cottin.c:2004}
C.~Cottin-Bizonne, C.~Barentin, E.~Charlaix, L.~Bocquet and J.~L. Barrat,
  \emph{Eur. Phys. J. E}, 2004, \textbf{15}, 427.

\bibitem{zhou.j:2012}
J.~J. Zhou, A.~V. Belyaev, F.~Schmid and O.~I. Vinogradova, \emph{J. Chem.
  Phys.}, 2012, \textbf{136}, 194706.

\bibitem{philip.jr:1972}
J.~R. Philip, \emph{J. Appl. Math. Phys.}, 1972, \textbf{23}, 353--372.

\bibitem{hocking.lm:1976}
L.~M. Hocking, \emph{J. Fluid Mech.}, 1976, \textbf{76}, 801--817.

\bibitem{wang2004}
C.~Wang, \emph{Physics of Fluids}, 2004, \textbf{16}, 2136.

\bibitem{priezjev.nv:2006}
N.~V. Priezjev and S.~M. Troian, \emph{J. Fluid Mech.}, 2006, \textbf{554},
  25--46.

\bibitem{niavarani.a:2010}
A.~Niavarani and N.~V. Priezjev, \emph{Phys. Rev. E}, 2010, \textbf{81},
  011606.

\bibitem{hendy2005effect}
S.~C. Hendy, M.~Jasperse and J.~Burnell, \emph{Physical Review E}, 2005,
  \textbf{72}, 016303.

\bibitem{bib:succi-01}
S.~Succi, \emph{The lattice {B}oltzmann equation for fluid dynamics and
  beyond}, Oxford University Press, 2001.

\bibitem{bib:bgk}
P.~L. Bhatnagar, E.~P. Gross and M.~Krook, \emph{Phys. Rev.}, 1954,
  \textbf{94}, 511.

\bibitem{bib:jens-kunert-herrmann:2005}
J.~Harting, C.~Kunert and H.~Herrmann, \emph{Europhys. Lett.}, 2006,  328--334.

\bibitem{bib:jens-jari:2008}
J.~Hyv\"aluoma and J.~Harting, \emph{Phys. Rev. Lett.}, 2008, \textbf{100},
  246001.

\bibitem{bib:zhu-tretheway-petzold-meinhart-2005}
L.~Zhu, D.~Tretheway, L.~Petzold and C.~Meinhart, \emph{J. Comp. Phys.}, 2005,
  \textbf{202}, 181.

\bibitem{bib:benzi-etal-06}
R.~Benzi, L.~Biferale, M.~Sbragaglia, S.~Succi and F.~Toschi, \emph{Europhys.
  Lett.}, 2006, \textbf{74}, 651.

\bibitem{bib:zhang-kwok-04}
J.~Zhang and D.~Y. Kwok, \emph{Phys. Rev. E}, 2004, \textbf{70}, 056701.

\bibitem{benzi-etal-06b}
R.~Benzi, L.~Biferale, M.~Sbragaglia, S.~Succi and F.~Toschi, \emph{Phys. Rev.
  E}, 2006, \textbf{74}, 021509.

\bibitem{bib:huang-thorne-schaap-sukop-2007}
H.~Huang, D.~T. Thorne, M.~G. Schaap and M.~C. Sukop, \emph{Phys. Rev. E},
  2007, \textbf{76}, 066701.

\bibitem{bib:jens-schmieschek:2010}
S.~Schmieschek and J.~Harting, \emph{Comm. Comp. Phys.}, 2011, \textbf{9},
  1165.

\bibitem{succi02}
S.~Succi, \emph{Phys. Rev. Lett.}, 2002, \textbf{89}, 064502.

\bibitem{bib:tang-tao-he-2005}
G.~H. Tang, W.~Q. Tao and Y.~L. He, \emph{Phys. Fluids}, 2005, \textbf{17},
  058101.

\bibitem{bib:sbragaglia-succi-2005}
M.~Sbragaglia and S.~Succi, \emph{Phys. Fluids}, 2005, \textbf{17}, 093602.

\bibitem{bib:ansumaili-karlin-2002}
S.~Ansumali and I.~V. Karlin, \emph{Phys. Rev. E}, 2002, \textbf{66}, 026311.

\bibitem{bib:sofonea-sekerka-2005}
V.~Sofonea and R.~F. Sekerka, \emph{Phys. Rev. E}, 2005, \textbf{71}, 066709.

\bibitem{bib:niu-shu-chew-2004}
X.~D. Niu, C.~Shu and Y.~T. Chew, \emph{Europhys. Lett.}, 2004, \textbf{67},
  600.

\bibitem{HechtHarting2010}
M.~Hecht and J.~Harting, \emph{Journal of Statistical Mechanics: Theory and
  Experiment}, 2010, \textbf{2010}, P01018.

\bibitem{ahmed.nk:2009}
N.~K. Ahmed and M.~Hecht, \emph{J.~Stat. Mech. - Theory and Exp.}, 2009,
  P09017.

\bibitem{alexeyev:96}
A.~A. Alexeyev and O.~I. Vinogradova, \emph{Colloids Surfaces A}, 1996,
  \textbf{108}, 173 -- 179.

\end{thebibliography}

\end{document}